\documentclass[reprint,
 showpacs,preprintnumbers,
 amsmath,amssymb,
 aps,
]{revtex4-1}

\usepackage{graphicx}
\usepackage{dcolumn}
\usepackage{bm}
\begin{document}

%\preprint{APS/123-QED}

\title{Effective dielectric response of
dispersions of graded particles}

\author{M.~Ya.~Sushko}
\email{mrs@onu.edu.ua} \affiliation{Department of Theoretical
Physics and Astronomy, Mechnikov National University, 2 Dvoryanska
St., Odesa 65026, Ukraine}

%\date{\today}

\begin{abstract}
Based upon our compact group approach and the Hashin-Shtrikman
variational theorem, we propose a new solution, which effectively
incorporates many-particle effects in concentrated systems,  to
the problem of the effective quasistatic permittivity of
dispersions of graded dielectric particles. After the theory is
shown to recover existing analytical results and simulation data
for dispersions of hard dielectric spheres with power-law
permittivity profiles, we use it to describe the effective
dielectric response of nonconducting polymer-ceramic composites
modeled as dispersions of dielectric core-shell particles.
Possible generalizations of the results are specified.
%\begin{description}
%\item[PACS numbers] 42.25.Dd, 77.22.Ch, 77.84.Lf, 82.70.-y
%\end{description}
\end{abstract}

 \maketitle

\section{\label{sec:intro} Introduction}

The effective properties of dispersions of graded particles have
been studied intensively. Besides being important due to their
abundance in nature \cite{Ohshima12}, such systems attract
particular interest because of their potential technical
applications. For instance, the effective dielectric response of
graded composites can be tuned finely by designing the dielectric
profiles of the constituent. Numerous applications of such systems
are discussed in
\cite{Bohren1983,Bergman1992,Nan1993,Garrou1998,Sihvola1999,Tsang2001,Torquato2002,Milton2002}.

Despite serious efforts made so far, the determination of the
effective dielectric properties of dispersions of graded
particles, which is the objective of the present report, remains a
challenging theoretical problem, even in the quasistatic limit.
The reason is strong variations of the electric field in such
systems due to electromagnetic interactions and spatial
correlations of their constituents. As a result, reliable
analytical results, such as \cite{Sihvola1989, Dong2003,Wei2003,
Wei2005} for graded spherical inclusions or
\cite{Gu2003,Gu2005,Wei2007} for two-dimensional composites of
graded cylindrical inclusions, are rare and have been obtained
within the Maxwell-Garnett approach
\cite{Maxwell1873,Maxwell1904,Landau1982} for diluted low-contrast
dispersions, where the interparticle influences are negligibly
weak: the problem is actually reduced to solving the governing
equation for the electric potential of a single particle in a
uniform external field \cite{Sihvola1989, Dong2003,Wei2003,
Wei2005,Gu2003,Gu2005,Wei2007}. Similarly, computer simulations
\cite{Mejdoubi2007NumericPermittivity} deal with the response of a
single graded particle in the computational domain. This situation
tacitly implies the use of the Maxwell-Garnett type of
homogenization.

The results \cite{Dong2003,Wei2003, Wei2005,Gu2003} validated
approximate schemes known as the differential effective dipole
approximation (DEDA) \cite{Yu2003,Huang2003}, anisotropic
differential effective dipole approximation (ADEDA)
\cite{Dong2004}, and differential effective multipole moment
approximation (DEMMA) \cite{Yu2003,Yu2005}. Based on the
application of differential analysis to available rigorous
solutions for a single graded particle, these schemes allowed one
to evaluate the induced dipole moments of single graded particles
with more complex shapes and, in general, anisotropic permittivity
profiles. The differential analysis also became an integral part
of other approaches, such as the differential replacement
procedure (DPR) \cite{Duan2006}, which combines it with an energy
equivalency condition; the multiscale homogenization scheme
\cite{Giordano2008}, proceeding from a generic theorem on the
equivalence between a graded dielectric ellipsoid and an
anisotropic homogeneous ellipsoid, etc.

It is significant to emphasize that after the individual
polarizability of a graded particle is found with the just-listed
or similar methods, the effective permittivity of a dispersion of
such particles is calculated using again the Maxwell-Garnett
approach. As the filler particle concentration or the
particle-matrix contrast or both are increased, the electric field
distribution in the dispersion becomes extremely complicated and
difficult to visualize. The pertinent microscopic calculations,
which now require the knowledge of an infinite set of the
correlation functions of the system, become virtually impossible.

By this report, we would like to attract the Reader's attention to
new developments of our earlier results
\cite{Sushko2007,Sushko2009CompGroups,Sushko2009AnisPart} and
their applications to dispersions of graded hard dielectric
spheres with piecewise-continuous permittivity profiles. They  are
obtained using our original method of compact groups of
inhomogeneities
\cite{Sushko2007,Sushko2009CompGroups,Sushko2009AnisPart}. It is
designed to effectively take into account many-particle
polarization and correlation effects in a system with complex
microstructure without uncontrolled assumptions about them. The
essential details  of the method are presented in
section~\ref{sec:summary}. The equation for the effective
dielectric constant of the dispersions is derived in
section~\ref{sec:eeff}. This equation still contains an unknown
parameter, the permittivity of the host in the auxiliary system;
combining our approach with the Hashin--Shtrikman variational
theorem \cite{Hashin1962}, this parameter is determined in
section~\ref{sec:hs}. The results of contrasting our theory with
other authors' analytical and numerical results are given in
section~\ref{sec:comparison}. The application of the theory to
nonconducting dispersions of core-shell particles are discussed in
section~\ref{sec:core-shell}. The main results obtained and
further suggestions are summarized in section~\ref{sec:concl}.

\section{\label{sec:summary} Summary of the compact group approach }
The main points of this approach are as follows.

(1) The effective quasistatic  permittivity
$\varepsilon_{\rm{eff}}$ of a system is defined by the relation
\cite{Landau1982}
\begin{equation} \label{effpermittivitydefinition}
\langle {\bf{D}} ({\bf{r}})\rangle = \langle \varepsilon
({\bf{r}}) {\bf{E}} ({\bf{r}}) \rangle = \varepsilon_{\rm{eff}}
\langle {\bf{E}} ({\bf{r}}) \rangle,
\end{equation}
where ${\bf{D}} ({\bf{r}})$, ${\bf{E}} ({\bf{r}})$, and
$\varepsilon ({\bf{r}})$  are the local values of,  respectively,
the electric induction, electric field, and permittivity in the
system. The angle brackets in (\ref{effpermittivitydefinition})
mean either statistical averaging or averaging by volume
integration; for an infinite system, both ways are expected to
give, according to the ergodic hypothesis
\cite{Landau1982,Torquato2002}, the same result.

(2) A dispersion $\cal{D}$ to be homogenized is equivalent, in its
long-wave dielectric response, to an auxiliary system $\cal{S}$
prepared by embedding the constituents (particles and matrix) of
$\cal{D}$ into a host (perhaps, imagined), $\cal{M}$, of some
permittivity $\varepsilon_{\rm f}$. The system $\cal{S}$ can be
viewed as a set of compact groups of particles and regions
occupied by the real matrix. The compact groups are defined as
macroscopic regions whose typical sizes are much smaller than the
wavelength $\lambda$ of probing radiation in $\cal{M}$, but which
yet include sufficiently large numbers $N$ of particles to remain
macroscopic and retain the properties of the entire $\cal{S}$. The
permittivity distribution in $\cal{S}$ is
\begin{equation}
\label{localpermittivity} \varepsilon ({\bf{r}}) =
\varepsilon_{\rm f}+ \delta\varepsilon ({\bf{r}}),
\end{equation} where $\delta\varepsilon
(\bf{r})$ is the contribution from a compact group located at
point $\bf{r}$. The explicit form of $\delta\varepsilon
({\bf{r}})$ is modeled in accord with the geometrical parameters
and dielectric properties of $\cal{D}$'s constituents.

(3) The analysis of the electric field distribution in a system
with permittivity distribution (\ref{localpermittivity}) is based
on the equation describing the propagation of an electromagnetic
wave in inhomogeneous media. After this equation is replaced by an
equivalent integral equation, the formal solutions for the
electric field and induction are represented in the form of
infinite iterative series. In the limit $\lambda\to \infty$,
fluctuations of particle numbers in compact groups and
correlations between the particles in different groups are
negligibly small, whereas the groups themselves are actually, with
respect to the probing radiation, one-point inhomogeneities. Their
contributions are formed by those ranges of coordinate values
where the electromagnetic field propagators reveal a singular
behavior. Decomposing each propagator into a Dirac delta function
part and a principal value part (see the end of section
\ref{sec:hs}), and using symmetry reasoning, the averaged
contributions from compact groups to $\langle {\bf{E}}\rangle$ and
$\langle {\bf{D}}\rangle$ can be singled out from all terms in the
iterative series for ${\bf{E}} ({\bf{r}})$ and ${\bf{D}}
({\bf{r}})$. These contributions are shown to dominate in the
formation of the long-wavelength dielectric characteristics of the
system.  As a result, for macroscopically homogeneous and
isotropic systems we have
\begin{equation} \label{averagefield}
\langle{\rm {\bf  {E}}}\rangle = {\left[ {1 + {\sum\limits_{s =
1}^{\infty}  {\left( { - {\frac{{1}}{{3\varepsilon _{\rm f}}} }}
\right)^{s} \langle {{\mathop {\left( {\delta \varepsilon ({\rm
{\bf r}})} \right)^{s}}}}} }}\rangle \right]}\,{\rm {\bf E}}_{0},
\end{equation}
\begin{equation}\label{averagedisplacement}
\langle{\rm {\bf {D}}}\rangle = \varepsilon _{\rm f}{\left[ {1 - 2
{\sum\limits_{s = 1}^{\infty}  {\left( { -
{\frac{{1}}{{3\varepsilon _{\rm f}}} }} \right)^{s}\langle
{{\mathop {\left( {\delta \varepsilon ({\rm {\bf r}})}
\right)^{s}}}}} }}\rangle  \right]}\,{\rm {\bf E}}_{0},
\end{equation}
where $\textbf{E}_{0}$ is the amplitude of the incident wave field
in ${\cal M}$. For models with bounded and piecewise-continuous
$\delta \varepsilon ({\rm {\bf r}})$, Eqs. (\ref{averagefield})
and (\ref{averagedisplacement}) are rigorous in the quasistatic
limit. In particular, together with
Eq.~(\ref{effpermittivitydefinition}) they reproduce the classical
result \cite{Landau1982}
\begin{equation*}\label{LandauLimit}
\varepsilon_{\rm eff}\approx\overline{\varepsilon}
-\frac{1}{3\overline{\varepsilon}}\overline{\left(\varepsilon-\overline{\varepsilon}\right)^2},
\end{equation*} which is valid, to $O\left(
\left(\varepsilon-\overline{\varepsilon}\right)^2/\overline{\varepsilon}^2\right)$,
for any mixture in which the local deviations of permittivity
$\varepsilon$ from the average value $\overline{\varepsilon}$
[$=\varepsilon_{\rm f} + \langle\delta \varepsilon ({\rm \bf
r})\rangle$] are weak.

Thus, the analysis of $\varepsilon _{\rm eff}$ reduces to modeling
$\delta \varepsilon ({\rm {\bf r}})$, calculating its moments
$\langle \left( {\delta \varepsilon ({\rm {\bf r}})} \right)^{s}
\rangle$, finding the sums in Eqs.~(\ref{averagefield}) and
(\ref{averagedisplacement}), and deciding on the value of
$\varepsilon _{\rm f}$.

\section{\label{sec:eeff} The equation for $\varepsilon_{\rm eff}$ of graded spheres}

For a dispersion of inhomogeneous hard dielectric spheres, with
radius $R$ and piecewise-continuous permittivity profile
$\varepsilon_1=\varepsilon_1({\bf r})$, embedded in a uniform
matrix, with constant permittivity $\varepsilon_0$,
\begin{eqnarray}
\delta \varepsilon ({\bf{r}}) =\Delta\varepsilon_0\left(1- \sum_{a
=1}^{N}  \theta (R - |{\bf{r}} - {\bf{r}}_{a}|)\right) \nonumber\\
\label{compgroupcontribution} +\sum_{a =1}^{N} \Delta\varepsilon_1
({\bf{r}} - {\bf{r}}_{a}) \theta (R - |{\bf{r}} - {\bf{r}}_{a}|),
\end{eqnarray}
where  $\Delta\varepsilon_0 = \varepsilon_0 -\varepsilon_{\rm f}$,
$\Delta\varepsilon_1({\bf{r}})= \varepsilon_1({\bf{r}})
-\varepsilon_{{\rm f}}$, $\theta(x)$ is the Heaviside step
function, and the summation is carried out over the position
vectors ${\bf{r}}_{a}$ of $N$ spheres belonging to the compact
group at ${\bf{r}}$.

For this $\delta \varepsilon ({\rm {\bf r}})$, both direct volume
integration \cite{Sushko2009CompGroups, Sushko2009AnisPart} and
statistical averaging \cite{Sushko2009JPS} give
\begin{equation}\label{moments}
{\langle {\mathop {\left( {\delta \varepsilon ({\rm {\bf r}})}
\right)^{s}}}\rangle}  = (1-c) \left( \Delta \varepsilon_0
\right)^{s} + n {\int\limits_{\Omega} {d{\rm {\bf r}}\, {\left(
\Delta \varepsilon_1 ({\rm {\bf{r}}}) \right)}^{s}}} , \quad s \ge
1,
\end{equation}
where $n =N/V$ and $c =4\pi R^3 n/3$ are the number density and
volume concentration of spheres, respectively, and the integral is
taken over the sphere volume $\Omega$. Then the sums in
Eqs.~(\ref{averagefield}) and (\ref{averagedisplacement}) are
found readily and together with
Eq.~(\ref{effpermittivitydefinition}) give
\cite{Sushko2009AnisPart}
\begin{equation}\label{effectivepermittivity}
(1-c)\frac{\varepsilon_0-\varepsilon_{\rm f}}{2\varepsilon_{\rm
f}+\varepsilon_0}+ n\int\limits_\Omega d{\bf r}
\frac{\varepsilon_1({\bf r})-\varepsilon_{\rm
f}}{2\varepsilon_{\rm f}+\varepsilon_1({\bf r})}=
\frac{\varepsilon_{\rm eff }-\varepsilon_{\rm
f}}{2\varepsilon_{\rm f}+\varepsilon_{\rm eff }}.
\end{equation}

In the case of isotropic spheres with radially-symmetric
permittivity profile $\varepsilon_1=\varepsilon_1(r)$,
Eq.~(\ref{effectivepermittivity}) can be represented as
\begin{equation}\label{effectivepermittivityradial}
(1-c)\frac{\varepsilon_0-\varepsilon_{\rm f}}{2\varepsilon_{\rm
f}+\varepsilon_0}+ 3 c\int\limits_0^1 du\,u^2
\frac{\varepsilon_1(u)-\varepsilon_{\rm f}}{2\varepsilon_{\rm
f}+\varepsilon_1(u)}= \frac{\varepsilon_{\rm eff
}-\varepsilon_{\rm f}}{2\varepsilon_{\rm f}+\varepsilon_{\rm eff
}},
\end{equation}
where $u=r/R$ is the dimensionless variable  and
$\varepsilon_1(u)$ is the sphere's permittivity profile as a
function of $u$.

For uniform spheres with $\varepsilon_1= {\rm const}$,
Eqs.~(\ref{effectivepermittivity}) and
(\ref{effectivepermittivityradial}) reduce to the classical
Maxwell-Garnett mixing rule \cite{Maxwell1873,Maxwell1904} if
$\varepsilon_{\rm f }=\varepsilon_0$, and the Bruggeman mixing
rule \cite{Bruggeman1935,Landauer1952,Stroud1975} if
$\varepsilon_{\rm f}=\varepsilon_{\rm eff}.$ These choices of
$\varepsilon_{\rm f}$ are known as the Maxwell-Garnett and the
Bruggeman types of homogenization, respectively. Which one must be
used in a particular case has been a matter of long-lasting
discussions (see, for instance,
\cite{Bohren1983,Bergman1992,Nan1993,Sihvola1999,Tsang2001,Torquato2002}).
To determine $\varepsilon_{\rm f}$ self-consistently, we use the
idea \cite{Sushko2016} to combine the compact group approach with
the Hashin-Shtrikman variational theorem \cite{Hashin1962} and
require that two different ways of homogenization -- through the
linear relation (\ref{effpermittivitydefinition})  and through the
equality of the electrostatic energies stored in $\cal{D}$ and
$\cal{S}$ -- give equal results.

Typically, theorem \cite{Hashin1962} is  used to determine the
most restrictive bounds for $\varepsilon_{\rm eff}$ of
macroscopically homogeneous and isotropic two-constituent
materials which can be derived in terms of the (definite)
constituent permittivities and volume concentrations. Attempts to
make use of it to evaluate these bounds for heterogeneous media
with graded constituents were made, for example, in
\cite{Wang2006,Duan2006} by introducing comparison materials with
microstructures different from those of the considered
heterogeneous media.

\section{\label{sec:hs} Hashin-Shtrikman theorem and the choice of $\varepsilon_{\rm f}$}

Let ${\bf E}_0$ and
\begin{equation}\label{HS5}
{\bf{D}}_0=\varepsilon_{\rm f} {\bf{E}}_0
\end{equation}
be the electric field and induction in ${\cal M}$ filling a large
region of volume $V$, provided the region's boundary ${\cal B}$ is
maintained at a prescribed time-independent potential $\psi ({\cal
B})=\psi_0 ({\cal B})$ and there are no free charges inside. Next,
suppose that the whole of the region is changed to a material with
permittivity (\ref{localpermittivity}), but without changing $\psi
({\cal B})$ and adding free charges. Then, according to the
Hashin-Shtrikman variational theorem \cite{Hashin1962}, the
functional (of ${\bf{T}} \equiv{\bf{D}}-\varepsilon_{\rm f}
{\bf{E}}$)
\begin{equation}\label{HS9}
U_{{\bf{T}}}=\frac{1}{8 \pi}\int\limits_V \left[\varepsilon_{\rm
f} {\bf{E}}^2_0\ -\frac{{\bf{T}}^2}{\varepsilon - \varepsilon_{\rm
f}} +2{\bf{T}}\cdot {\bf{E}}_0+ {\bf{T}}\cdot \left({\bf{E}} -
{\bf{E}}_0\right)\right] d{\bf r},
\end{equation}
subject to the subsidiary condition
\begin{equation}\label{HS10}
\varepsilon_{\rm f}\, \textrm{div} \left ( {\bf{E}} - {\bf{E}}_0
\right ) + \textrm{div}\, {\bf{T}}=0,
\end{equation}
is stationary for
\begin{equation}\label{HS12}
{\bf{T}} = \left( \varepsilon - \varepsilon_{\rm f}\right)
{\bf{E}},
\end{equation}
and its stationary value $U_{\textbf{T}}^{\rm s}$ is the
electrostatic energy stored in $V$.

Note that Eq.~(\ref{HS12}) is equivalent to
$${\bf{D}}({\bf{r}})=\varepsilon({\bf{r}}){\bf{E}}({\bf{r}}),$$
where $\varepsilon({\bf{r})} = \varepsilon_{\rm f} + \delta
\varepsilon({\bf{r}})$, and  Eq.~(\ref{HS10}) is fulfilled.

In view of Eq.~(\ref{HS12}), the integrand  in Eq.~(\ref{HS9})
reduces to $\varepsilon_{\rm f}
\textbf{E}^2_0+\left(\varepsilon-\varepsilon_{\rm f}
\right)\textbf{E} \textbf{E}_0$. Hence, using
Eqs.~(\ref{effpermittivitydefinition}) and (\ref{averagefield})
and denoting
\begin{equation}\label{matrix2}
Q\equiv\sum_{s=1}^{\infty}\left(-\frac{1}{3\varepsilon_{\rm f}}
\right)^s \left( \delta \varepsilon ({\bf{r}}) \right)^s,
\end{equation}
we find
\begin{equation}\label{matrix4}
U_{{\bf{T}}}^{\rm s} = \frac{V {\bf{E}}_0^2}{8\pi} \left[
\varepsilon_{\rm f}+\left(\varepsilon_{\rm eff} - \varepsilon_{\rm
f} \right) \left( 1+\langle Q \rangle \right) \right].
\end{equation}
It is natural to require that this value be equal to the
electrostatic energy stored in the homogenized system. Then, in
view of Eqs.~\eqref{effpermittivitydefinition} and
\eqref{averagefield}, we also have
\begin{equation}\label{matrix7}
U_{{\bf{T}}}^{\rm s}=\frac{V}{8\pi} \langle {\bf{E}} \rangle
\langle {\bf{D}} \rangle= \frac{V{\bf{E}}_0^2}{8\pi}
\,\varepsilon_{\rm eff} \left(1+ \langle Q \rangle\right)^2.
\end{equation}
Finally,  Eqs.~(\ref{effpermittivitydefinition}),
(\ref{averagefield}), and (\ref{averagedisplacement}) yield
\begin{equation}\label{matrix5}
\langle {\bf{D}} \rangle=  \varepsilon_{\rm f} \left(1-2 \langle Q
\rangle \right) {\bf{E}}_0 =\varepsilon_{\rm eff} \left(1+ \langle
Q \rangle\right) {\bf{E}}_0.
\end{equation}

Eqs.~(\ref{matrix4}), (\ref{matrix7}), and (\ref{matrix5}) give a
system of homogeneous linear equations in $\varepsilon_{\rm f}$
and $\varepsilon_{\rm eff}$:
\begin{equation}\label{matrix8}
\varepsilon_{\rm f}+\left(\varepsilon_{\rm eff} - \varepsilon_{\rm
f} \right) \left( 1+\langle Q \rangle \right) = \varepsilon_{\rm
eff} \left(1+ \langle Q \rangle\right)^2,
\end{equation}
\begin{equation}\label{matrix9}
\varepsilon_{\rm f}\left({1-2\langle Q
\rangle}\right)=\varepsilon_{\rm eff} \left({1+\langle Q
\rangle}\right).
\end{equation}
The nontrivial solutions exist provided
\begin{equation}\label{matrix10}
\langle Q \rangle(1-\langle Q \rangle)=0.
\end{equation}

In the case
\begin{equation}\label{matrix11}
\langle Q \rangle=0
\end{equation}
$\varepsilon_{\rm f}=\varepsilon_{\rm eff}$. If  $\langle Q
\rangle=1$, then  $\varepsilon_{\rm f}=-2\varepsilon_{\rm eff}$.
The latter situation may occur for metamaterials
\cite{Veselago1967,Mackay2004}, but is beyond the scope of the
present work.

It follows that within the compact group approach supplemented by
the Hashin-Shtrikman theorem \cite{Hashin1962}, both ways of
homogenization (through the linear relation between the average
induction and average field, and through the equality of the
electrostatic energies of the heterogeneous and homogenized
systems) are consistent provided the Bruggeman-type homogenization
$\varepsilon_{\rm f}=\varepsilon_{\rm eff}$ is used. The
looked-for $\varepsilon_{\rm eff}$ is the solution to
Eq.~(\ref{matrix11}). No special assumptions about the composition
of the system, the geometry and concentration of the constituents,
and the permittivity distribution (except for its piecewise
continuity) in the system were used in the above discussion.

It should be emphasized that our approach is not equivalent to the
classical Bruggeman mean-field approximation \cite{Bruggeman1935}
(and for this reason, the term "Bruggeman-type homogenization" is
used for the result $\varepsilon_{\rm f}=\varepsilon_{\rm eff}$).
The latter approximation is traditionally understood as a
one-particle approach where a single particle is placed in the
effective medium of permittivity $\varepsilon_{\rm eff}$  (see
\cite{Bergman1992}). In contrast, our theory effectively
incorporates many-particle effects, and the result
$\varepsilon_{\rm f}=\varepsilon_{\rm eff}$   is derived for the
situation where a macroscopically-large (compact) group of
particles is placed in the effective medium of permittivity
$\varepsilon_{\rm eff}$. It is the effects of interparticle
polarizations and correlations in such groups that determine the
behavior of $\varepsilon_{\rm eff}$  in the limit $\lambda\to
\infty$.

Note that Eq.~(\ref{matrix11}) can be represented as
\begin{equation}\label{KerRelation}
\left\langle \frac{\varepsilon ({\bf r})-\varepsilon_{\rm
f}}{2\varepsilon_{\rm f}+\varepsilon ({\bf r})} \right\rangle=0.
\end{equation}
This is exactly the condition $\langle\xi ({\bf r})\rangle=0$
imposed on the stochastic field $ \xi ({\bf r})= (\varepsilon
({\bf r})-\varepsilon_{\rm f})/(2\varepsilon_{\rm f}+\varepsilon
({\bf r}))$ in  the strong-property-fluctuation theory (SPFT)
\cite{Tsang2001,Ryzhov1965,Ryzhov1970,Tamoikin1971,Tsang1981,Zhuck1994,Michel1995,Mackay2000,Mackay2001}
in order to improve the convergence of the iteration procedure
applied to an integral equation for $\bf E({\bf r})$. The
resulting Dyson-type equation for $\langle{\bf E}({\bf r})\rangle$
is usually analyzed using the second-order truncation of the mass
operator series (bilocal approximation
\cite{Tsang2001,Ryzhov1970,Zhuck1994,Michel1995,Mackay2000}; on
the significance of the third-order approximation, see
\cite{Mackay2001}), the Gaussian statistics for $\xi ({\bf r})$,
and model expressions for the two-point correlation function
$G({\bf r},{\bf r^\prime})=\langle\xi({\bf r})\xi({\bf
r^\prime})\rangle$. The symmetry of $G({\bf r},{\bf r^\prime})$
dictates the shape of the exclusion volume required for the
decomposition of the propagator into a principal value part and a
Dirac delta function part.

Contrastingly, in the compact group approach we use, for the
propagators in the iterative series for $\textbf{E}(\textbf{r})$
and $\textbf{D}(\textbf{r})$, the decomposition with  a spherical
exclusion volume, in conformity with the requirement that both the
entire system and compact groups be macroscopically homogeneous
and isotropic. The statistical microstructure of the system comes
into play only at the final stage,  as the moments $\langle \left(
\delta \varepsilon (\textbf{r}) \right)^s \rangle$ are calculated.

For dispersions of spheres with permittivity profiles
$\varepsilon_1 = \varepsilon_1 ({\bf r})$ and $\varepsilon_1 =
\varepsilon_1 ({ r})$, Eq.~(\ref{matrix11}) reduces to,
respectively,  Eqs.~(\ref{effectivepermittivity}) and
(\ref{effectivepermittivityradial}) at $\varepsilon_{\rm
f}=\varepsilon_{\rm eff}$ (then their right sides vanish).

\section{\label{sec:comparison} Comparison with other theories and simulations}

For low values of volume concentration $c$ and dielectric contrast
$\varepsilon_1/\varepsilon_0$, when both electromagnetic
interactions and spatial correlations are small, the
Maxwell-Garnett- and Bruggeman-type approaches are expected to
give close results. This gives a good reason  to contrast our
theory  with that \cite{Sihvola1989} for mixtures of hard spheres
with continuous radial permittivity profiles. Two ways were used
there to calculate the polarizability of small inhomogeneous
spheres in a quasistatic field and then $\varepsilon_{\rm eff}$:
the internal field method to find the dipole moment by integrating
the product of the field and the permittivity over the sphere's
volume, and the external field method to determine the field
perturbation due to the sphere and then the amplitude of the
equivalent dipole. Both ways lead to the same results, found by
solving differential equations analytically; therefore, results
\cite{Sihvola1989} are free of ambiguities typical of computer
simulations using the finite-difference method (see, for instance,
\cite{Karkainen2001}).

\begin{figure}
\centering
\includegraphics[width=75mm]{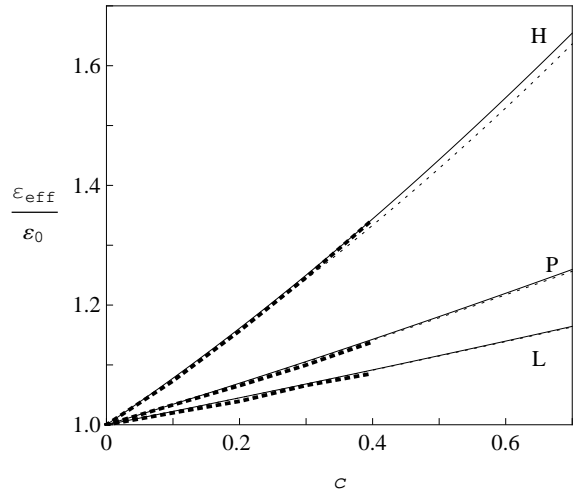}
\caption{{ $\varepsilon_{\rm eff}/\varepsilon_0$ versus volume
concentration  $c$ for dispersions of inhomogeneous hard spheres
with the H-, L-, and P-types of the permittivity profiles (see
text) according to Eq.~(\ref{effectivepermittivityradial}) with
$\varepsilon_{\rm f}=\varepsilon_{\rm eff}$ (solid lines) and
$\varepsilon_{\rm f}=\varepsilon_0$ (thin dotted lines). The thick
dotted lines represent analytical results \cite{Sihvola1989}.}}
\label{fig:CompSihvola1}
\end{figure}

Figure~\ref{fig:CompSihvola1} represents  our results obtained
with Eq.~(\ref{effectivepermittivityradial}) for $\varepsilon_{\rm
eff}$ of dispersions of hard dielectric spheres embedded in a
medium of permittivity $\varepsilon_0=1$. The spheres have the
following permittivity profiles ($0 \leq u \leq 1$): homogeneous
$\varepsilon_1(u){= 2\, \varepsilon_0}$ (denoted as H); linear
$\varepsilon_1(u)= \varepsilon_0 (2-u)$ (L); parabolic
$\varepsilon_1(u)= \varepsilon_0 \left(2-u^2\right)$ (P).  The
agreement of our results with analytical results
\cite{Sihvola1989} turns out to be surprisingly good in the entire
interval $c\in [0,0.4]$, investigated in \cite{Sihvola1989}. It
follows that our theory works well even in the situation where the
concept of compact groups may seem questionable.

\begin{figure}
\centering
\includegraphics[width=75mm]{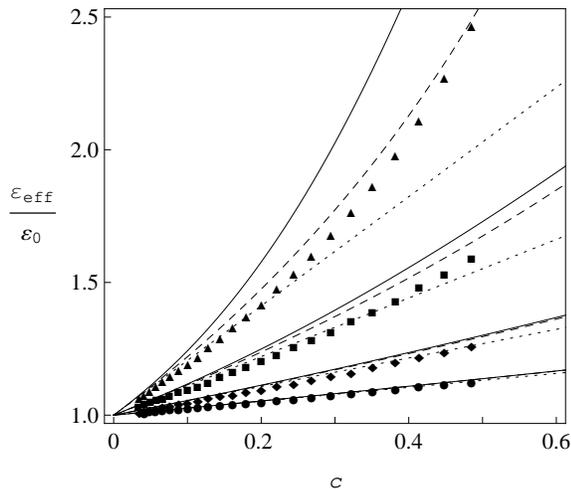}
\caption{{ $\varepsilon_{\rm eff}/\varepsilon_0$ versus volume
concentration $c$ for dispersions of hard dielectric spheres with
power-law permittivity profiles $\varepsilon_1(u)= (2+u)^k$ at $k=
0.25, \,0.5,\,1$, and 2 according to: final-element modeling
\cite{Mejdoubi2007NumericPermittivity} ($\bullet$,\,
$\blacklozenge$,\, $\blacksquare$,\, and $\blacktriangle$,
respectively); Eq.~(14) in \cite{Wei2005} (dotted lines);
Eq.~(\ref{effectivepermittivityradial}) for $\varepsilon_{\rm
f}=\varepsilon_{\rm eff}$ (solid lines) and $\varepsilon_{\rm
f}=\varepsilon_0$ (dashed lines).}} \label{fig:CompWeiMejdoubi}
\end{figure}

This fact is further confirmed by the comparison of our theory
with analytical  \cite{Wei2003,Wei2005} and simulation
\cite{Mejdoubi2007NumericPermittivity} results for hard dielectric
spheres with the power-law permittivity profiles
$\varepsilon_1(u)= c(b+u)^k$; those results are shown in
Fig.~\ref{fig:CompWeiMejdoubi} for $c=1$, $b=2$, and $k= 0.25,
\,0.5,\,1,\,2$. In \cite{Wei2003,Wei2005}, the local electrical
potentials for isolated spheres were derived rigorously in terms
of the hyper-geometric function; then they were used to predict,
based on the Maxwell-Garnett approach, the effective dielectric
response of the graded composites in the dilute limit. The authors
of \cite{Mejdoubi2007NumericPermittivity} reported their effective
permittivity calculations with the finite-element method for
two-phase graded composite materials. The agreement between
predictions \cite{Wei2005} and simulation data
\cite{Mejdoubi2007NumericPermittivity} is sufficiently good only
for low-contrast dispersions ($c=1$, $b=2$, and $k= 0.25, \,0.5$),
or in the dilute limit. As Fig.~\ref{fig:CompWeiMejdoubi} reveals,
our theory reproduces these results as well. For higher contrasts
$\varepsilon_1/\varepsilon_0$ ($k=1$ and 2), the discrepancies
between the theories become considerable. Our predictions for
$\varepsilon_{\rm eff}$, given by
Eq.~(\ref{effectivepermittivityradial}) with $\varepsilon_{\rm
f}=\varepsilon_{\rm eff}$,  are always greater than those reported
in \cite{Wei2003,Wei2005, Mejdoubi2007NumericPermittivity}. This
can be explained by the fact that our theory effectively takes
into account multiparticle effects, whereas the studies
\cite{Wei2003,Wei2005, Mejdoubi2007NumericPermittivity} are
actually one-particles approaches. It is interesting to note that
for all above $k$'s, the formal use of the Maxwel-Garnett type of
homogenization within our formalism
[Eq.~(\ref{effectivepermittivityradial}) with $\varepsilon_{\rm
f}=\varepsilon_0$] gives results that are close to simulation data
\cite{Mejdoubi2007NumericPermittivity}.

\section{\label{sec:core-shell} Nonconducting dispersions of core-shell particles}

One of the practical applications of our theory is the prediction
of the dielectric properties of microelectronic and optoelectronic
devices requiring, for their superior performance, the use of
packing materials with low dielectric constants, low dielectric
losses, and high volume resistivities \cite{Garrou1998}. Since
packaging materials are often polymer-ceramic composites, numerous
core-shell theories have been developed (see, for instance,
\cite{Xue2000,Vo2002,Todd2003,Sang2005,Liu2010,Liu2011} and
literature therein) in which $\varepsilon_{\rm eff }$ of a
composite is characterized by the geometric and permittivity
parameters of the polymer phase, the filler phase, and an
interphase region within the composite system. The equation for
$\varepsilon_{\rm eff}$ is usually derived in these theories in
several steps: (1) combining the particle and the adjacent
interphase layer into a ``complex particle''; (2) finding the
effective polarizability of the isolated complex particle in a
uniform field; (3) calculating the effective dielectric constat of
the system within the standard one-particle  approaches
\cite{Maxwell1873,Maxwell1904,Bruggeman1935,Landauer1952,Landau1982}
or their modifications. In contrast, we use only step (1) and then
calculate $\varepsilon_{\rm eff}$ treating the system in terms of
compact groups of such complex particles.

Suppose that each filler particle, of radius $R$ and permittivity
$\varepsilon_1$, is surrounded by a concentric particle-matrix
interphase shell, of inner radius $R$, outer radius $R+t$, and
permittivity $\varepsilon_2$. Considering the particle and the
adjacent interphase shell to be a single hard particle, we readily
find from Eq.~(\ref{effectivepermittivityradial}) the equation for
$\varepsilon_{\rm eff}$ of dispersions obtained by embedding such
particles into a matrix of permittivity $\varepsilon_0$:
\begin{eqnarray}
\left[1-\phi(c,\delta)\right]\frac{\varepsilon_0-\varepsilon_{\rm
eff}}{2\varepsilon_{\rm eff}+\varepsilon_0}+
c\frac{\varepsilon_1-\varepsilon_{\rm eff}}{2\varepsilon_{\rm
eff}+\varepsilon_1} \nonumber \\
\label{effectivepermittivitycoreshell}+
\left[\phi(c,\delta)-c\right] \frac{\varepsilon_2-\varepsilon_{\rm
eff}}{2\varepsilon_{\rm eff}+\varepsilon_2}=0.
\end{eqnarray}
Here, $\delta=t/R$ is the relative thickness of the shell, and
$\phi(c,\delta)$ is the effective volume concentration of complex
particles (that is, the sum of the volume concentration $c$ of the
filler and that of the interphase region). For hard filler
particles,
\begin{equation} \label{effconchard}
\phi(c,\delta)=(1+\delta)^3c.
\end{equation}

Note that Eq.~(\ref{effectivepermittivitycoreshell}), but with
different $\phi(c,\delta)$, also holds in the case where the
interphase shells can be treated as fully penetrable
(freely-overlapping) \cite{Sushko2013} (see also
\cite{Tomylko2015}). For such systems, the scaled-particle
estimate \cite{Rikvold85} gives
\begin{eqnarray}\phi(c,\delta)= 1-
(1 - c)\,\exp\left[{-\frac{((1+\delta)^3 -
1)c}{1-c}}\right] \nonumber\\
 \times  \exp\left\{- \frac{3(1 + \delta)^3
c^2}{2(1 - c)^3} \left[2 - \frac{3}{1+\delta} +
\frac{1}{(1+\delta)^3} \right. \right. \nonumber \\
\label{effconc1} - \left. \left. \left( \frac{3}{1+\delta} -
\frac{6}{(1+\delta)^2} + \frac{3}{(1+\delta)^3}\right) c
\right]\right\}.
\end{eqnarray}
As $\delta\to 0$, $\phi(c,\delta)=
(1+3\delta+3\delta^2)c+O(c\delta^3)$ and tends to the value
(\ref{effconchard}).

\begin{figure}
\centering
\includegraphics[width=75mm]{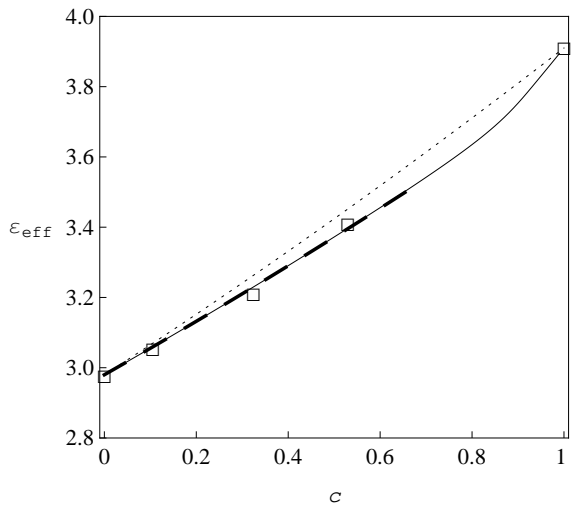}
\caption{{ Experimental data  \cite{Todd2003} ($\square$) for
$\varepsilon_{\rm eff}$ of polymer-ceramic composite samples of
near-spherical amorphous ${\rm SiO_2}$ particles embedded randomly
in an epoxy matrix, and their fits with: the Bruggeman equation
[in Eq.~(\ref{effectivepermittivitycoreshell}), $\delta=0$ and
${\phi(c,0)=c}\,$] for two-constituent composites (dotted line);
Eqs.~(\ref{effectivepermittivitycoreshell}) and
(\ref{effconchard}) (dashed line);
Eqs.~(\ref{effectivepermittivitycoreshell}) and (\ref{effconc1})
(solid line). The matrix permittivity $\varepsilon_0=2.98$ and the
${\rm SiO_2}$ particle permittivity $\varepsilon_1 =3.91$ were
extracted from Fig.~2 in \cite{Todd2003} as $\varepsilon_{\rm
eff}$ at $c=0$ and $c=1$, respectively; the shell permittivity
$\varepsilon_2 =2.5$ and relative thickness $\delta =0.06$ were
estimated by fitting.}} \label{fig:CompTodd}
\end{figure}

Figure~\ref{fig:CompTodd} represents the results obtained with
Eq.~(\ref{effectivepermittivitycoreshell}) for $\varepsilon_{\rm
eff}$ of model epoxy-based polymer-ceramic composites
\cite{Todd2003}, considered as dispersions of dielectric
core-shell spheres. When processing the data, we took into account
that: (1) the dielectric properties of constituents can change in
the composite preparation process; (2) homogenization theories are
intended to provide justified functional relationships between
$\varepsilon_{\rm eff}$ and the parameters of the actual
microstructure units in a real composite. For equal-sized spheres
with $R=4.5 \,{\rm \mu m}$ and $t=270\,{\rm nm}$ \cite{Todd2003},
the relative thickness $\delta = 0.06$. As is seen from
Fig.~\ref{fig:CompTodd}, our theory exactly recovers this value of
$\delta$. It also predicts a drop, to $\varepsilon_2=2.5$, in the
interphase permittivity, as compared to the matrix permittivity.
This estimate differs from that given by theory \cite{Vo2002},
$\varepsilon_2=2.88$, by about 15\%. Physically,  the drop in
$\varepsilon_2$ can be attributed \cite{Todd2003} to the chemical
bonding of the polymer to the filler particle surface.

It should be remarked that the relation of the parameter $\delta$,
calculated with the model of equal-sized spheres, to the actual
thickness $d$ of the interphase in a real dispersion can be rather
complicated due to the distribution of particle diameters and
shapes. In particular,  applying our general results
\cite{Sushko2009AnisPart} to mixtures of hard core-shell particles
(not necessarily spherical) with different core sizes $R_a$ and
fixed interphase thickness $d \ll R_a$, it is readily to show that
$\varepsilon_{\rm eff}$ of such mixtures is given by
Eq.~(\ref{effectivepermittivitycoreshell}), but with $\phi
=c(1+S_{\rm p}d/V_{\rm p})$ instead of (\ref{effconchard}). Here
$V_{\rm p}$ and  $S_{\rm p}$ are the total volume and the total
surface area of the particles, respectively. Correspondingly,
$$d\approx \left[(1+\delta)^3-1\right]V_{\rm p}/S_{\rm p}.$$ A
convincing example in favor of this relation was given in
\cite{Todd2003}: after the actual particle size distribution of
${\rm Si O_2}$ particles was taken into account, the estimate $d
\approx 5\, {\rm nm}$ was obtained. The latter is consistent with
available literature data (see \cite{Todd2003} for references).

\begin{figure}
\centering
\includegraphics[width=75mm]{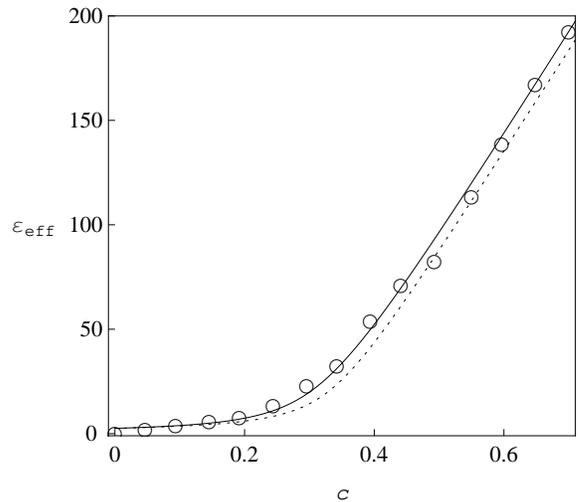}
\caption{{Experimental data \cite{Wang1996} ($\circ$) for
$\varepsilon_{\rm eff}$ of polystyrene-barium titanate composites
and their fits with: the Bruggeman equation
[Eq.~(\ref{effectivepermittivitycoreshell}) at $\delta=0$ and
$\phi(c,0)=c$] for two-constituent composites (dotted line);
Eqs.~(\ref{effectivepermittivitycoreshell}) and
(\ref{effconchard}) for hard core-shell particles (solid line).
The matrix permittivity was taken to be $\varepsilon_0=2.55$. The
permittivity  of $\rm BaTiO_3$ particles, that of the interphase
shells, and the relative thickness of the shells were estimated at
$\varepsilon_1=330$, $\varepsilon_2=19$, and $\delta =0.09$,
respectively.}} \label{fig:CompWang}
\end{figure}

To finish, we also test the applicability of our theory to
high-contrast systems, such  as polystyrene-barium titanate
composites \cite{Wang1996} (see also \cite{Xue2000}). Experimental
data \cite{Wang1996} are widely-cited, but the parameters used to
process them with existing theories and the results obtained
differ considerably. For example, for $\varepsilon_0=2.55$, some
reported values of $\varepsilon_1$ are 250 \cite{Todd2005,Liu2011}
and 800 \cite{Xue2000,Liu2010}, those of $\varepsilon_2$ are 3
\cite{Xue2000}, 7.24 \cite{Liu2011}, and 25 \cite{Liu2010}, and
those of $\delta$ are 0.005 \cite{Xue2000}, 0.025 [recovered from
Eq.~(6) in \cite{Liu2011} at $k=0.161$], and 0.26 \cite{Liu2010}.
Our estimates (see Fig.~\ref{fig:CompWang}) fall in these ranges.
This means that the functional structure of
Eq.~(\ref{effectivepermittivitycoreshell}) is capable of
reproducing available experimental data. However, an in-depth
analysis of possible factors (size and shape distributions,
dielectric losses, inhomogeneities of the interphase, etc.) behind
the above discrepancies is beyond the scope of the present report.

\section{\label{sec:concl} Conclusion}
The main results of this report can be summarized as follows.

(1) Combining the compact group approach \cite {Sushko2007,
Sushko2009CompGroups, Sushko2009AnisPart} with the
Hashin-Shtrikman variational theorem \cite{Hashin1962} and
requiring that the two common ways for homogenization (through the
linear material equation between the induction and the field, and
through the equality of the electric energies of the heterogeneous
and homogenized systems)  give equal results, we proposed a new
solution, which effectively incorporates many-particle effects in
concentrated systems,  to the problem of the effective quasistatic
permittivity $\varepsilon_{\rm eff}$ of dispersions of graded
dielectric particles.  According to it: a dispersion to be
homogenized is dielectrically equivalent to a macroscopically
homogeneous and isotropic system prepared by embedding the
constituents of the real dispersion into an imagined medium having
the looked-for permittivity (Bruggeman-type homogenization); the
equation for $\varepsilon_{\rm eff}$ is an integral relation
obtained from Eq.~(\ref{matrix11}) by summing up the statistical
moments for the local deviations of the permittivity distribution
in the model system from $\varepsilon_{\rm eff}$; this equation
validates the condition postulated for the relevant stochastic
field in the SPFT. The latter fact is a strong argument for our
theory because the SPFT has proved to be very efficient for
homogenization of composites of non-graded constituents.

(2) The efficiency of our theory was demonstrated by contrasting
its results with analytical results
\cite{Sihvola1989,Wei2003,Wei2005} and simulation data
\cite{Mejdoubi2007NumericPermittivity} for dispersions of hard
dielectric spheres with power-law permittivity profiles. The
theory was also applied to nonconducting polymer-ceramic
composites considered as dispersions of dielectric core-shell
spheres. The comparison of its results with experimental data
\cite{Xue2000,Todd2003,Wang1996} showed that the theory can be
used to predict the effective dielectric response of such systems
in terms of the geometric and dielectric parameters of their
constituents, including the interphase regions.

The generalizations of these results are possible in two
directions, at least. First, the compact group approach is
expected to be applicable to multiconstituent dispersions of
anisotropic inhomogeneous dielectric particles, as long as these
dispersions remain macroscopically homogeneous and isotropic. Some
relevant general results for $\varepsilon_{\rm eff}$ of such
systems are presented in \cite{Sushko2009AnisPart}. Second, this
approach can be extended to dispersions of particles with complex
permittivities. In particular, it has already been shown to be
efficient for the description of electric percolation phenomena in
composites of core-shell particles \cite{Sushko2013}, two-step
electrical percolation in nematic liquid crystals filled with
multiwalled carbon nanotubes \cite{Tomylko2015}, and the effective
structure parameters of suspensions of nanosized insulating
particles \cite{Sushko2016JML}.

\bigskip

\begin{acknowledgments}
I thank Prof. A. Lakhtakia and  Dr. T. G.  Mackay  for providing
copies of their articles and A. V. Dorosh for checking some
graphical materials.
\end{acknowledgments}

\end{document}